%% file: acl_latex.tex
\pdfoutput=1
\documentclass[11pt]{article}

\usepackage{acl}

\usepackage{times}
\usepackage{latexsym}

\usepackage[T1]{fontenc}
\usepackage[utf8]{inputenc}

\usepackage{microtype}

\usepackage{inconsolata}
\usepackage[normalem]{ulem}
\usepackage{amsmath}
\usepackage{amssymb}
\usepackage{amstext}
\usepackage{bm}
\usepackage{bbm}
\usepackage{algorithm}
\usepackage{algpseudocode}
\usepackage{amsmath}
\usepackage{multirow}
\usepackage{graphicx}
\usepackage{subfigure}
\usepackage{epsfig}
\usepackage{float}
\usepackage{enumitem}
\usepackage{epstopdf}
\usepackage{framed}
\usepackage{url}
\usepackage{stfloats}
\usepackage{xspace}
\usepackage{booktabs}
\usepackage{subeqnarray}
\usepackage{enumerate}
\usepackage{color}
\usepackage{listings}
\usepackage{xcolor}

\newcommand{\paratitle}[1]{\vspace{1.5ex}\noindent\textbf{#1}}
\newcommand{\ie}{\emph{i.e.,}\xspace}

\newcommand{\eg}{\emph{e.g.,}\xspace}

\newcommand{\ignore}[1]{}

\usepackage{makecell}

\usepackage{cite}

\usepackage{CJKutf8}
\usepackage[utf8]{inputenc}

\title{BASES: Large-scale Web Search User Simulation with Large Language Model based Agents}

\author{\textbf{Ruiyang Ren\textsuperscript{\rm{1}}\thanks{~~The work was done during the internship at Baidu.} \quad 
Peng Qiu\textsuperscript{\rm{2}} \quad
Yingqi Qu\textsuperscript{\rm{2}} \quad
Jing Liu\textsuperscript{\rm{2}}\thanks{~~Corresponding authors.} \quad 
Wayne Xin Zhao\textsuperscript{\rm{1}}}\footnotemark[2] \\ 
\textbf{Hua Wu\textsuperscript{\rm{2}} \quad 
Ji-Rong Wen\textsuperscript{\rm{1,3}} \quad  
Haifeng Wang\textsuperscript{\rm{2}}  
}\\
	\textsuperscript{1}Gaoling School of Artificial Intelligence, Renmin University of China  \\
	\textsuperscript{2}Baidu Inc. \\
        \textsuperscript{3}School of Information, Renmin University of China  \\
	\{reyon.ren, jrwen\}@ruc.edu.cn, batmanfly@gmail.com\\
	\{qiupeng02, quyingqi, liujing46, wu\_hua, wanghaifeng\}@baidu.com
}

\begin{document}
\maketitle
\begin{abstract}
Due to the excellent capacities of large language models (LLMs), it becomes feasible to develop LLM-based agents for reliable user simulation. Considering the scarcity and limit (\eg privacy issues) of real user data, in this paper, we conduct large-scale user simulation for web search, to improve the analysis and modeling of user search behavior.   
Specially, we propose BASES, a novel user simulation framework with LLM-based agents, designed to facilitate comprehensive simulations of web search user behaviors. 
Our simulation framework can generate unique user profiles at scale, which subsequently leads to diverse search behaviors.  
To demonstrate the effectiveness of BASES, we conduct evaluation experiments based on two human benchmarks in both Chinese and English,  demonstrating that BASES can effectively simulate large-scale human-like search behaviors.
To further accommodate the research on web search, we develop WARRIORS, a new large-scale dataset encompassing web search user behaviors, including both Chinese and English versions, which can greatly bolster research in the field of information retrieval.
Our code and data will be publicly released soon. 
\end{abstract}

\input{sec/intro}

\input{sec/related.tex}

\input{sec/simulation.tex}

\input{sec/application.tex}

\input{sec/dataset.tex}

\input{sec/conclusions.tex}

\section*{Ethical Considerations}
Given the limitations and privacy concerns associated with real user data, we introduce BASES, a framework for simulating large-scale web search user behaviors using LLM-based agents. BASES demonstrates promising potential in enhancing information retrieval tasks, especially in low-resource scenarios. This approach not only respects user privacy but also opens up new avenues for research in this domain, contributing positively to the advancement of information retrieval technologies. As we move forward, we are committed to ensuring that our simulation practices uphold the ethical standards and contribute constructively to the field.

\section*{Limitations}

Our user simulation methodology exhibits highly realistic simulated behaviors and good application performance. However, due to the complexity of human, the user profiles we constructed do not comprehensively encapsulate all human characteristics. Although minor, this fact also harbors the potential to influence search behaviors of users. In the future, we will continue to explore schemes for simulating web search users, aiming to align simulated behaviors as closely as possible with those of real users.

% \section*{Acknowledgements}

% Bibliography entries for the entire Anthology, followed by custom entries
%\bibliography{anthology,custom}
% Custom bibliography entries only
\bibliography{custom}

\newpage

\appendix

\input{sec/appendix}

% This is an appendix.

\end{document}

%% file: sec/intro.tex
\section{Introduction}

Web search is a typical information-seeking scenario, where a user issues a query, retrieves the web pages from the search engine, and subsequently selects and browses through the pages of interest. 
The core to the success of search systems lies in the accurate understanding and modeling of user behavior.  
Thus, by analyzing large-scale user behavior under web search scenarios, we can enhance the understanding of user's information needs, thereby facilitating the development of more effective search systems~\citep{DBLP:conf/sigir/BennettWCDBBC12, DBLP:conf/sigir/ZhouDWXW21, DBLP:conf/cikm/Chen0MZSMZM21}.
However, existing research works typically rely on real-user experiments, which can be costly~\citep{DBLP:conf/trec/CarteretteKHC14}. Moreover, concerns regarding the quality and integrity of collected user data emerge as notable impediments, potentially undermining the precision of the analysis~\citep{DBLP:conf/www/SugiyamaHY04}. 
In addition, these experiments encompass ethical considerations, such as user privacy issues. {As a result, user behavior simulation emerges as a promising direction~\citep{DBLP:conf/cikm/JiangHHYN12}.}
 
Recent years have witnessed the unprecedented success of large language models~(LLMs)~\citep{DBLP:journals/corr/abs-2303-18223}. With superb model capabilities, LLMs are capable of comprehending complex instructions and executing actions as \emph{autonomous agents}~\citep{DBLP:journals/corr/abs-2309-07864, DBLP:journals/corr/abs-2308-11432}.  
Prior studies have explored LLM-based agents in various fields, such as recommender systems~\citep{DBLP:journals/corr/abs-2306-02552} and {dialogue systems}~\citep{DBLP:journals/corr/abs-2308-07201}.
In contrast with prior studies, we focus on a quintessential user interaction scenario of web search and utilize LLM-based agents to simulate web search user behaviors. 
Despite that this idea is intuitive, there exist several major challenges in simulating the  behaviors of web search users. 
Firstly, each user is an independent individual, it is difficult to ensure that each user has a unique and reasonable profile, particularly at a large scale.
Secondly, predicting web search user behaviors with precision and personalization is challenging.

\begin{figure}
    \centering
    \includegraphics[width=0.49\textwidth]{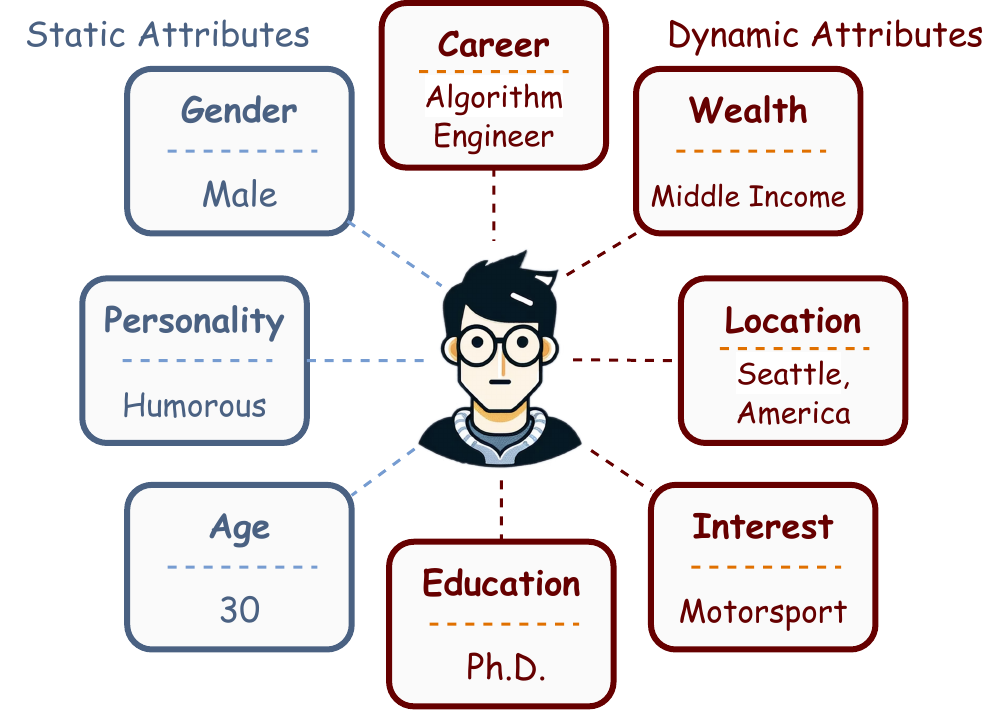}
    \caption{The profile structure for web search user simulation with a user sample.}
    \label{fig:profile}
\end{figure}

To address these issues, we propose a novel framework \textbf{BASES} for large-scale we\underline{B} se\underline{A}rch u\underline{SE}r \underline{S}imulation with LLM-based agents.
First, we design user profile attributes tailored to the characteristics of web search users with reference to the related study~\citep{DBLP:conf/adaptive/2007}. Each user profile encapsulates two basic categories, static and dynamic, and a total of eight attributes are {allocated} within the two categories. Figure~\ref{fig:profile} gives an example of a user profile.
To construct unique profiles for large-scale simulated users,
we propose the synergistic synthesis method, which predefines a comprehensive range of potential attribute values combining manual definition with the cooperation of GPT-4, ensuring both efficiency and diversity for user profile generation. 
Subsequently, LLM-based agents tailored with distinct user profiles are employed to conduct user simulations through the proposed query and click behavior prompting strategies in the search engine, thereby generating precise and personalized user behaviors.

To verify the effectiveness of the BASES framework, we first 
show the high anthropomorphism in search behaviors of the LLM-based agents by quantitative analysis, and conduct
the manual evaluation on the simulated behavior data to demonstrate the personalization degree. 
Following these foundational assessments, we adopt BASES within both Chinese and English settings on two information retrieval~(IR) tasks, including session search and click prediction.
We find BASES-based models achieve an improvement of up to 13\% in NDCG@1 compared to other behavior datasets trained models, while utilizing less than one percent or one thousandth of their training data volume.
Moreover, BASES also shows considerable potential in the adaption of low-resource scenarios, bringing up to 7\% improvements in NDCG@1 with multiple adapting strategies.
Finally, to accommodate the progressively evolving search patterns of users, we systematically collect the simulated data and construct a new dataset WARRIORS of \underline{W}eb se\underline{AR}ch use\underline{R} behav\underline{IORS} with 100,000 users' search sessions for both Chinese and English versions. 
The key contributions are summarized as follows:
\begin{itemize}[topsep=2pt,parsep=0pt,partopsep=0pt,leftmargin=1.2em]
    \item We propose BASES, a novel web search user behavior simulation framework based on LLM-based agents, involving constructing extensive and diverse user profiles and precisely performing user behaviors in web search scenarios. 
    \item Extensive experiments on the proposed BASES framework across various IR tasks demonstrate that BASES significantly enhances the performance of IR models, and operates effectively in low-resource scenarios.
    \item We systematically collect and organize the large-scale simulated web search user behaviors, and release a new dataset WARRIORS, encompassing web search behaviors in both Chinese and English versions.

\end{itemize}

%% file: sec/related.tex
\section{Related Work}

\paratitle{Behavior Analysis for Web Search.}
Web search behavior analysis~\citep{DBLP:journals/cn/HolscherS00} represents a crucial research direction within the realm of information retrieval, focusing on tasks such as session search and click models. 
Session search particularly focuses on understanding the information needs of users throughout their behavior sessions~\citep{DBLP:journals/jasis/JansenSBK07}. This includes grasping how users' search intents evolve over time and how to predict future behaviors~\citep{DBLP:conf/sigir/JiangHA14, DBLP:conf/cikm/0001NDMZDZJ21, DBLP:conf/sigir/ZhangMLMZM20}, with several benchmarks proposed to support studies of session search~\citep{DBLP:conf/cikm/ChenMLZM19, DBLP:conf/trec/CarteretteKHC14}.
In the research of session search, click models hold a central position, including probabilistic graphical models
~\citep{DBLP:conf/www/ChapelleZ09}, user browsing models~\citep{DBLP:conf/sigir/DupretP08}, sequential models~\citep{DBLP:conf/sigir/BorisovWMR18}, and machine learning models~\citep{DBLP:conf/wsdm/ChenML0M20}. They are designed to capture behavior patterns by analyzing extensive query-click data~\citep{DBLP:series/synthesis/2015Chuklin}. Through these techniques, a more comprehensive understanding of search behaviors can be attained, leading to the creation of search engines that are more effective and closely tailored to user needs.

\paratitle{User Simulation.}
User simulation emerges as a pivotal research area that offers a cost-effective approach for approximating behaviors of real users~\citep{DBLP:books/daglib/0088953}, thereby addressing the prevalent issues of scarcity and incompleteness in real data.
Continuous research efforts have been directed toward user simulation over the years and have made a series of progress~\citep{DBLP:conf/interspeech/GeorgilaHL06, DBLP:journals/taslp/SchatzmannY09, DBLP:journals/ijhci/BiswasRL12}.
Recent advancements in LLMs~\citep{DBLP:journals/corr/abs-2303-18223} significantly propel the field of user simulation and explore across multiple domains, encompassing dialogue generation~\citep{DBLP:journals/corr/abs-2308-07201}, recommendation systems~\citep{DBLP:journals/corr/abs-2306-02552, DBLP:journals/corr/abs-2310-09233}, car driving~\citep{DBLP:journals/corr/abs-2309-13193}, and social behavior~\citep{DBLP:conf/uist/ParkOCMLB23, DBLP:journals/corr/abs-2209-06899}. This marks a significant departure from traditional user simulation methodologies, offering more nuanced and contextually rich user interaction models. 
Our study contributes to this evolving field by focusing on web search, a classic but under-explored area, leveraging LLM-based agents to perform large-scale user simulations, demonstrating a broad spectrum of practical applications.

%% file: sec/simulation.tex
\section{Large-scale User Simulation for Web Search}
\label{sec:simulation}
In this section, we propose a novel framework for large-scale user \underline{Sim}ulation in \underline{W}eb \underline{S}earch scenarios based on LLM-based agents, called \textbf{BASES}, with specially designed  user profile construction strategies and a tailored simulation process. We also demonstrate the  feasibility of the simulation framework by evaluating the human-agent consistency in terms of their search behavior patterns.

\subsection{User Profile Construction} 
In this part, we detail the process of user profile construction, including user profile structure, construction methods for user profiles, and the definition of attribute values.

\subsubsection{User Profile Structure}
\label{sec:profile_structure}

To effectively simulate the  search behaviors of web users, it is crucial to establish diverse and detailed user profiles tailored to web search scenarios. The profile should encompass various attributes that exert varying degrees of influence on search behaviors. 
For example, a younger user may exhibit a preference for searching and clicking on more trendy topics.
Drawing on insights from prior research work~\citep{DBLP:conf/adaptive/2007, DBLP:conf/fskd/Liang11}, as shown in Figure~\ref{fig:profile}, we design a profile structure of web search users, comprised of eight distinct attributes, including age, gender, education, career, personality, interest, location, and wealth, which are further categorized into static attributes and dynamic attributes. 
Static attributes are inherent to the user and tend to remain constant over time~(\eg gender), while dynamic attributes reflect transient and external aspects and are subject to change~(\eg interest). 
It is essential for each simulated web search user to possess a unique profile, thereby enabling the simulation of a diverse range of search behaviors.

\subsubsection{User Profile Construction Methods}
\label{sec:profile_methods}
To construct diverse and authentic user profiles, a basic method is involving human participants in filling out profile attributes. Despite its flexibility, it is inefficient and limited by the human annotator. 
Another intuitive method is utilizing LLMs such as GPT-4~\citep{DBLP:journals/corr/abs-2303-08774} with instructions to automatically generate user profiles, but it is hard to ensure diversity.

To this end, we consider two profile construction methods: 
\uline{\textit{Real Data based Generation}} analyzes real user data through LLMs to generate the user profile. This method requires real user data and may be affected by the data quality.
\uline{\textit{Synergistic Synthesis}} scientifically predefines attribute values through a collaborative effort involving human and GPT-4. These predefined attribute values are then randomly sampled based on various well-designed strategies, such as distribution patterns and logical coherence. 
Given the requirement to simulate web search users at a large scale, where both efficiency and diversity are paramount in profile construction, we adopt synergistic synthesis as the primary profile construction method.

With synergistic synthesis for user profile construction, we employ three principles for predefining attribute value candidates and sampling: (1) For attributes with easily-defined and uniformly-distributed candidate values (\eg gender), we manually define each attribute value and conduct random sampling. (2) For attributes with unclear value candidates and even distribution (\eg interest), we employ a coarse-to-fine sampling approach. First, we categorize attributes into various coarse-grained categories, subsequently integrating GPT-4 for the generation of fine-grained attribute values with manual adjustment within each category. (3) For attributes exhibiting non-uniform distributions of value candidates (\eg location), we consult pertinent literatures to define potential attribute values, sampling them in accordance with their distribution. 
In addition, we meticulously identify and address the irrational aspects in profiles, including age-career contradictions, age-education contradictions, and career-education contradictions.

\subsubsection{Attribute Value Definition}

Here, we detail the specific definition and sampling methods employed in each attribute.

\noindent\textbf{Age:}
In consideration of the authentic demographics of web search users, we implement random selections of user ages ranging from 6 to 90. 

\noindent\textbf{Gender:}
We implement random sampling to select user gender, including male and female.

\noindent\textbf{Personality:}
We initially delineate the coarse-grained personality categories referring to the five-factor model~\citep{mccrae1992introduction}, including openness, conscientiousness, extraversion, agreeableness, and neuroticism. Subsequently, we define the fine-grained personality values with the collaboration of human and GPT-4 in each category.

\noindent\textbf{Education:}
Current statistics regarding the global population's educational levels are somewhat incomplete.
Nevertheless, we refer to the reports from the UNESCO Institute for Statistics~\citep{education} and the World Bank~\citep{worldbank} to approximately estimate the distribution of different educational levels of users.

\noindent\textbf{Career:}
Referring to 
ISIC\footnote{\url{https://unstats.un.org/unsd/classifications/Econ/isic}} from the United Nations, we initially delineate twenty coarse-grained career categories such as ``Financial and insurance activities'', student category and Non-working attributes are also included. Subsequently, we define the fine-grained career values with the collaboration of human and GPT-4 in each category.

\noindent\textbf{Interest:}
Collaborating with GPT-4, we initially delineate eleven coarse-grained interest categories such as ``Sports and outdoor activities''. Subsequently, we define the fine-grained interest values with the collaboration of human and GPT-4 within each coarse-grained category.

\noindent\textbf{Location:}
We consult global demographic data\footnote{\url{https://worldpopulationreview.com/world-cities}} in 2023 to identify the top 1,000 cities ranked by population in English setting, serving as candidate locations. For the Chinese setting, we use cities in China with populations
as candidates~\citep{ning2021main}.
Then, we randomly sample cities weighted according to the population distribution of the cities.

\noindent\textbf{Wealth:}
We categorize users into three distinct wealth status groups: low income, middle income, and high income. 
Referring to the existing studies~\citep{wealth}, 
we randomly sample each wealth category weighted by corresponding approximate population proportion.

\subsection{Web Search User Simulation}
Based on the aforementioned profile construction method, we employ LLM-based agents in search engine environments to conduct large-scale  user behavior simulation for web search. 

\subsubsection{Simulation Essentials}

In our BASES framework, the agents simulating users are required to be aware of their current state in the search scenario, plan to proactively formulate search queries, assimilate information returned by search engines, and selectively click web pages that match their interests. Additionally, the agents need to reflect whether their information needs have been satisfied based on the retrieved information, to decide whether new queries are warranted for further searches. 

Although existing studies have successfully utilized  LLM-based agents for  simulating and fulfilling particular tasks under different environments~\citep{DBLP:conf/uist/ParkOCMLB23, DBLP:journals/corr/abs-2306-02552} , the capacity of LLM-based agents to accurately mimic real users in web search scenarios remains unclear.
We assess the consistency of behavior patterns between the agents and real users in the search engine to prove the feasibility of web search user simulation.
We find that LLM-based agents show about 90\% accuracies for query behaviors consistency and click behaviors consistency.
The details of the assessment can be found in Appendix~\ref{appendix:consistency}.

\subsubsection{Simulation Process}

\paratitle{Behavior Patterns.} We employ the web search user profiles constructed using synergistic synthesis described in Section~\ref{sec:profile_structure}, each user profile corresponding to an LLM-based agent for user simulation. We utilize the ChatGPT with \texttt{GPT-3.5-Turbo-1106} version as the LLM to implement the web  user agents. These agents, acting on behalf of their respective users, engage in multi-round interactions in search sessions with search engines. We define three types of actions: search, click, and finish. In each round, an agent first generates a query based on its user profile and historical behaviors, which is then submitted to the search engine and the search engine returns relevant web pages from the internet. Subsequently, the agent selects web pages of interest to click on, {based on} its profile, historical behavior, the current query, and the returned web pages. At the end of each round, the agent decides whether to continue with another round or to finish the current session.

\paratitle{Behavior Prompting Strategies.} An intuitive idea for simulating user behavior with LLM-based agents, similar to ReAct ~\citep{DBLP:conf/iclr/YaoZYDSN023}, involves defining all possible actions within one instruction template, planning the overall steps before task execution, or thinking progressively during the task execution. 
{We make an attempt in a web search scenario} and find that predefining all possible actions within one instruction is not beneficial for the agents to effectively mimic human user behavior. {This issue is primarily manifested in the arbitrary click behaviors of the agents.}
Consequently, we devise two instruction strategies to prompt agents to execute precise actions, including query behavior prompting and click behavior prompting. 
\underline{\textit{Query behavior prompting}} is tailored to enable the agents to generate queries for web search based on their user profiles and historical behaviors. Considering the search pattern of real users, {we limit the queries to be concise}, primarily focusing on keyword keyphrases. The decision to finish or continue the session is also integrated, since the search action is the first action of each round.
If the agent opts to continue the session, it performs a search action to start a new round, otherwise, it executes a finish action. 
\underline{\textit{Click behavior prompting}} is designed to guide the agent in selecting web pages to view from search engine results, considering its user profile, current query, and historical behaviors. We carefully consider real web search scenarios, providing agents with the top ten web pages returned by search engines. When clicking, the agent primarily relies on the web page titles, and not only selects the webpage IDs but also provides explanations for their choices, so as to prevent arbitrary clicks.

For various prompting strategies, we design specific templates to ensure the outputs of agents are stable and controllable.
The full instruction templates are reported  in Appendix~\ref{appendix:prompts}.
Finally, each user exhibits sequential behaviors (\ie a search session) within the web search scenario.

\subsection{Discussion}

With our BASES framework, it is feasible to simulate highly personalized, diverse search behaviors via LLM-based agents. We propose profile construction methods capable of effortlessly generating tens of thousands of distinct user profiles, and efficiently and effectively producing the behaviors for these users through two behavior prompting strategies.
It greatly benefits the application scenarios where real user data is limited or difficult to collect.
Moreover, given the controllability of the simulated web search users, our BASES framework can flexibly incorporate various constraints to meet the further demands of specific tasks, such as conducting long search sessions, formulating queries in unique formats, and supporting distinct rules for clicks, among others.

Personalization is critical in web search behavior simulation, its absence can lead to homogenization in user behaviors, hindering subsequent tasks. 
{We manually assess the personalization in the simulated user behaviors of BASES, and find that even users with extremely similar profiles exhibited significantly inconsistent behavior patterns.} The details can be found in Appendix~\ref{appendix:personalization}.

%% file: sec/application.tex
\section{Effectiveness Verification}
In this section, we investigate the  effectiveness and applicability of our proposed BASES framework on classical information retrieval~(IR) tasks using evaluation benchmarks with real user data.

\subsection{Web Search User Simulation for IR tasks}
We first give necessary setups, then evaluate BASES on two real web search user benchmarks.

\subsubsection{Setups}
\label{sec:setup}

\paratitle{Task Formulation.}
In general, the BASES framework can be applied to various IR tasks, among which we consider two classical IR tasks: session search and click prediction. \textit{Session search} leverages the rich contextual information (such as queries $\{q_i\}_{i=1}^{n-1}$ and click history $\{d_i\}_{i=1}^{n-1}$) from earlier rounds of a user's session to aid document ranking for the subsequent search query $q_n$. It aims to enhance the learning of user preferences, thereby predicting the document $d_n$ most aligned with the user's intent from candidate documents $\{d_n^j\}_{j=1}^{k}$. The task of \textit{click prediction} aims to predict the documents most relevant to the current search query $q$ {from the search results $\{d_j\}_{j=1}^{k}$.}

\paratitle{Evaluation Model.}
We employ a BERT-based ranking model for evaluation on one NVIDIA GeForce RTX 3090, training up to 3 epochs with a batchsize of 128 and a learning rate of 5e-6. 
For the session search task, the model's input is the concatenation of historical user behavior sequence $\{q_i, d_i, ..., q_{n-1}, d_{n-1}\}$, the current query $q_n$, and the candidate document $d_n^j$. The output of the model is a relevance score between the document $d_n^j$ and the current query with behavior history. For the click prediction task, the model's input is the concatenation of query $q$ and candidate document $d_j$, with the output being the relevance score between the query $q$ and the candidate document $d_j$.

\paratitle{Evaluation Metrics.}
The evaluation metrics utilized in our study include Mean reciprocal rank~(MRR) and Normalized Discounted cumulative gain (NDCG). MRR quantifies search accuracy by averaging the reciprocal ranks of the first relevant result across queries. NDCG assesses ranking quality by comparing the weighted relevance of all results to an ideal order, and we use NDCG@1 and NDCG@3 for evaluation.

\begin{table*}[]
    \small
    \renewcommand\tabcolsep{4.3pt}
    \centering
    \begin{tabular}{clccccccc}
        \toprule
        \multirow{2.5}{*}{\textbf{Tasks}}&\multirow{2.5}{*}{\textbf{Methods}} & \multirow{1.5}{*}{\textbf{\#Session}} & \multicolumn{3}{c}{\textit{\textbf{Chinese Benchmark}}} & \multicolumn{3}{c}{\textit{\textbf{English Benchmark}}} \\
        \cmidrule(lr){4-6} \cmidrule(lr){7-9}
        & & (\multirow{0.5}{*}{\textbf{\#Click}}) & \textbf{MRR} & \textbf{NDCG@1} & \textbf{NDCG@3} & \textbf{MRR} & \textbf{NDCG@1} & \textbf{NDCG@3} \\
        \midrule
        \multirow{5}{*}{\textbf{Session}} & BM25 & - & 45.16 & 27.20 & 41.39 & 33.06 & 14.43 & 25.39   \\
        \multirow{5}{*}{\textbf{Search}}& BERT~(TREC-Session) & 1,257 & - & - & - & 32.85 & 13.11  & 25.89 \\
        & BERT~(AOL) & 219,748 & - & - & - & 35.59 & 16.07 & 29.54 \\
        & BERT~(Tiangong-ST) & 143,155 & 43.28 & 22.59 & 38.91 & - & - & - \\
        & BERT~(BASES) & 1,000 & \underline{51.78} & \underline{35.56} & \underline{47.98} & \underline{39.59} & \underline{19.02} & \underline{34.22} \\
        & BERT~(BASES) & 10,000 & \textbf{53.52} & \textbf{35.98} & \textbf{50.72} & \textbf{40.86} & \textbf{20.33} & \textbf{35.50} \\
        \midrule
        \multirow{5}{*}{\textbf{Click}}& BM25 & - & 42.71 & 21.34 & 39.96 & 31.35 & 11.80 & 23.88   \\
        \multirow{5}{*}{\textbf{Prediction}}& BERT~(TREC-Session) & 1,654 & - & - & - & 32.32 & 12.13 & 25.89 \\
        & BERT~(AOL) & 614,651 & - & - & - & 33.71 & 15.41 & 26.94 \\
        & BERT~(Tiangong-ST) & 318,823  & 41.29 & 21.34 & 36.28 & - & - & - \\
        & BERT~(BASES) & 2,880/2,920 & \underline{46.24} & \underline{27.20} & \underline{42.39} & \underline{36.74} & \underline{16.07} & \underline{30.37} \\
        & BERT~(BASES) & 28,511/28,802 & \textbf{49.11}  & \textbf{30.13} & \textbf{45.92} & \textbf{38.97} & \textbf{18.69} & \textbf{32.13} \\
        \bottomrule
    \end{tabular}
    \caption{Results of methods trained on different datasets in real user  benchmarks across two IR tasks.}
    \label{tab:benchmark}
\end{table*}

\subsubsection{Evaluation on Real User Behavior Benchmarks}
To demonstrate the efficacy of BASES, we employ it to generate behavior data of simulated web search users {in both Chinese and English settings} and train user behavior models for evaluation. We collect real user behaviors and construct benchmarks on the two classic IR tasks for evaluation.

\paratitle{Evaluation Benchmark.}
{In the Chinese setting}, we randomly sampled 100 anonymous sessions from {recent} search logs obtained via Baidu Search, with each session encompassing the user's historical queries, search results, and click records. {Since recent English user search behavior data is not available}, we develop an English benchmark through the collaboration of GPT-4 and humans. Initially, we instruct GPT-4 to produce query sequences similar to those generated by real users 
and obtain a collection of 100 diverse query sequences. Human annotators are then involved in clicking one of the search results considering history behavior and current query to {construct the search sessions.}

\paratitle{Baselines.}
For comparison, we select three widely adopted session search datasets for model training, including TREC-Session~\citep{DBLP:conf/trec/CarteretteKHC14} and AOL~\citep{verbeek2006user} for the English benchmark, and Tiangong-ST~\citep{DBLP:conf/cikm/ChenMLZM19} for the Chinese benchmark. We train BERT-based ranking models~(as described in Section~\ref{sec:setup}) on the three datasets respectively.

\paratitle{Results and Analysis.}
Table~\ref{tab:benchmark} reports the results of various methods on Chinese and English benchmarks constructed with real user behaviors. We observe that the models trained using user behavior data constructed by BASES significantly outperform the baselines in both session search and click prediction tasks, even with a small amount of training data. The results prove that BASES can accurately simulate web search users performing search processes in search engines like real users. 
Furthermore, we find that the classic lexical model BM25 shows considerable competitiveness, surpassing some semantic models trained on web search behavior datasets. As the performance of search engines has improved in recent years, enabling them to handle more complex queries. This has gradually influenced the query formats made by web search users, leading to discrepancies from the query formats in previous datasets.

\begin{table}[]
    \renewcommand\tabcolsep{3.2pt}
    \centering
    \small
    \begin{tabular}{lcccc}
        \toprule
        \textbf{Training Data} & \textbf{\#Session} & \textbf{MRR} & \textbf{NDCG@1} & \textbf{NDCG@3} \\
        \midrule
        Origin & 800 & 60.89 & 40.65 & 53.03 \\
        \midrule
        Ori. + Desc. & 1600 & \textbf{65.28} & \textbf{47.97} & 57.27 \\
        Ori. + Behav. & 1600 & 65.20 & 46.34 & \textbf{57.44} \\
        Ori. + Synth. & 1800 & 62.80 & 45.53 & 52.86 \\
        Ori. + Synth. & 2300 & 62.37 & 43.09 & 54.27 \\
        Ori. + Synth. & 2800 & 63.85 & 46.34 & 55.00 \\
        \bottomrule
    \end{tabular}
    \caption{Results of BASES-augmented methods on TREC-Session. \textit{Origin}~(\textit{Ori.}) denotes the training data of TREC-Session. \textit{Desc.}, \textit{Behav.}, and \textit{Synth.} denote the training data constructed by BASES with profiles generated by descriptions and user behaviors in TREC-Session, and synergistic synthesis, respectively. }
    \label{tab:low_resource}
\end{table}

\subsection{Adaptation on Low-resource Scenarios}

Given the inherent challenges in acquiring large-scale real user behavior data, the data scarcity poses great challenges for IR models.  Our framework BASES can generate simulated user behavior data, thus potentially improving the performance in low-resource web search scenarios.

\paratitle{Experimental Setup.}
We consider TREC-Session for experiments. {Since it is annotated by human participants, the overall volume of data is not substantial, which is a typical low-resource web search scenario}. In this scenario, we apply our BASES framework to generate augmented data on the session search task. We construct user profiles for simulation using methods in Section~\ref{sec:profile_methods}. The user profiles are generated with ChatGPT using real data based generation by employing descriptions (\textbf{Desc.}) or user behaviors (\textbf{Behav.}) in the dataset, and using the method of synergistic synthesis~(\textbf{Synth.}) without referring to the dataset.  
We also conduct experiments {with various scales of profiles generated by Synth..}
Subsequently, we utilize the three kinds of profiles to simulate augmented user behaviors for training. 
We use the same BERT-based ranking models for each setting.

\paratitle{Results and Analysis.}
Table~\ref{tab:low_resource} presents the results on TREC-Session employing various augmentation methods with BASES. It can be observed that all the methods with augmented data can yield improvements over the baseline established by training on the original TREC-Session training set. This demonstrates the effectiveness of the BASES framework to address the data scarcity issue in low-resource web search scenarios, thereby augmenting the original behavioral data to secure improved performance.
Furthermore, {augmentations with Synth. achieve improvements}, with the degree of improvement exhibiting an ascending trend in correlation with the amount of augmented data. 
{This phenomenon emphasizes the ability of BASES to bring substantial benefits in low-resource scenarios without necessitating dataset-specific information.
}

%% file: sec/dataset.tex
\begin{table}[t!]
    \centering
    \small
    \begin{tabular}{|c|c|c|}
        \hline
        \textbf{Dataset Statics} & \textbf{Chinese} & \textbf{English} \\
        \hline
        {\#user} & 100,000 & 100,000  \\
        \hline
        {\#avg. query per user} & 2.85  & 2.88   \\
        \hline
        {\#avg. click per query} & 1 &  1\\
        \hline
        {Search engine} & \multicolumn{2}{c|}{Google Search} \\
        \hline
        {Time}  & \multicolumn{2}{c|}{Feb., 2024} \\
        \hline
    \end{tabular}
    \caption{Statistics of the WARRIORS dataset.}
    \label{tab:datasets}
\end{table}

\begin{table*}[]
    \centering
    \small
    \begin{tabular}{llcccccc}
        \toprule
        \multirow{2.5}{*}{\textbf{Tasks}}&\multirow{2.5}{*}{\textbf{Methods}} & \multicolumn{3}{c}{\textit{\textbf{WARRIORS-Chinese}}} & \multicolumn{3}{c}{\textit{\textbf{WARRIORS-English}}} \\
        \cmidrule(lr){3-5} \cmidrule(lr){6-8}
        & & \textbf{MRR} & \textbf{NDCG@1} & \textbf{NDCG@3} & \textbf{MRR} & \textbf{NDCG@1} & \textbf{NDCG@3} \\
        \midrule
        \multirow{6}{*}{\textbf{Session Search}}& BM25 & 31.15 & 12.15 & 23.56 & 31.98 & 11.94 & 25.11 \\
        & BERT~(TREC-Session) & - & - & - & 33.33 & 13.78 & 26.60 \\
        & BERT~(AOL) & - & - & - & 34.85 & 14.70 & 28.24 \\
        & BERT~(Tiangong-ST) & 34.40 & 13.91 & 27.88  & - & - & - \\
        & BERT~(WARRIORS) & \textbf{59.40} & \textbf{39.73} & 58.97 & 46.96 & 27.29 & 43.32 \\
        & COCA~(WARRIORS) & 59.11 & 39.20 & \textbf{58.99} & \textbf{47.07} & \textbf{27.78} & \textbf{43.39}\\
        \midrule
        \multirow{5}{*}{\textbf{Click Prediction}}& BM25 & 30.44 & 11.62 & 23.09 & 32.10 & 12.12 & 25.20\\
        & BERT~(TREC-Session) & - & - & - & 34.03 & 13.93 & 27.62 \\
        & BERT~(AOL) & - & - & - & 31.94 & 11.84 & 24.79 \\
        & BERT~(Tiangong-ST) & 34.85 & 15.57 & 28.31 & - & - & - \\
        & BERT~(WARRIORS) & \textbf{57.49} & \textbf{38.01} & \textbf{56.73} & \textbf{46.68} & \textbf{27.00} & \textbf{43.20} \\
        \bottomrule
    \end{tabular}
    \caption{Evaluation of several baselines on WARRIORS-Chinese and WARRIORS-English.}
    \label{tab:baseline}
\end{table*}

\section{The WARRIORS Dataset}
% \textcolor{blue}{add transition and context} 
Previous experiments verified the effectiveness of the BASES framework. To this end, we construct \textbf{WARRIORS}, a large-scale simulation dataset of \underline{W}eb se\underline{AR}ch use\underline{R} behav\underline{IORS} based on BASES.

\subsection{Existing Issues}
For web search tasks, it is crucial to construct high-quality user behavior datasets. 
In the literature, several datasets have been proposed and contribute significantly to the advancement of the field. 
However, these datasets come with certain limitations. For instance, the AOL dataset~\citep{DBLP:conf/infoscale/PassCT06} is 18 years old now, where many URLs are no longer accessible. TREC-Session~\citep{DBLP:conf/trec/CarteretteKHC14} is constrained by crowdsourced experiments, lacking large-scale records of user behavior. Tiangong-ST~\citep{DBLP:conf/cikm/ChenMLZM19} is a large-scale web search behavior dataset in Chinese, defining session boundaries based on natural pauses of 30 minutes in user activity, which may introduce errors in session demarcation. 

Furthermore, web search user behaviors are highly affected by the underlying  algorithms of search engines. As the performance of search engines has gradually improved, the characteristics reflected in user behaviors within search engines have also evolved. For example, in the era of search engines reliant on keyword matching, users typically conducted searches using a single word or very short phrases. However, in recent years, with the rapid development of semantic search technologies~\citep{zhao2024dense}, user queries have shifted towards longer phrases and even sentences, as these approaches enable more efficient retrieval of the desired information. This shift implies that historical datasets of web search user behaviors may no longer reflect current user behaviors, presenting challenges for applying these datasets to enhance contemporary web search experiences.

Given the capability of BASES, it can be naturally used to construct a new dataset of web search behaviors using the data generated by {the anthropomorphic users simulated by LLM-based agents}.

\subsection{Construction of WARRIORS}

The construction of the WARRIORS dataset follows the BASES framework with LLM-based agents introduced in Section~\ref{sec:simulation}. The WARRIORS dataset is comprised of two versions: WARRIORS-Chinese and WARRIORS-English. For both versions, we use Google Search as the search engine. The user profiles of the two versions are independently generated, without overlapping users. 

In search engines, {the user behavior is characterized by the search session.
Each user session} encompasses multiple interaction rounds, each comprising a query, search results from the query, and clicked web pages from search results. The search results include the top ten web pages returned by the search engine,  each web page with its URL, title, and snippet. 
{Appropriate constraints are increased on the instructions for LLM agents, such as limiting the number of interaction rounds to avoid extremely long sessions.}   
The entire dataset is further divided into  training set,  validation set, and  test set. 
{Given the substantial volume of the dataset,} the specific allocation was 98\% for the training set, with the validation and test sets each receiving 1\%.
Table~\ref{tab:datasets} shows the details of the dataset.

\subsection{{Dataset Application}}
\paratitle{Experiment Setup.} 
WARRIORS can be easily applied to a variety of IR tasks, such as session search, click prediction, query suggestion, and so on. 
Here, we implement and evaluate several baselines on session search and click prediction tasks with both WARRIORS-Chinese and WARRIORS-English. For each task, we apply two representative models, a lexical matching model BM25~\citep{DBLP:conf/sigir/LinMLYPN21} and a BERT-based~\citep{DBLP:conf/naacl/DevlinCLT19} semantic ranking model. 
Furthermore, we report the results of a BERT-based ranking model trained using existing web search user behavior datasets~\citep{DBLP:conf/cikm/ChenMLZM19, DBLP:conf/trec/CarteretteKHC14} for reference.

\paratitle{Results and Analysis.}
Table~\ref{tab:baseline} reports the results of the baselines on WARRIORS-Chinese and WARRIORS-English. we find that the BERT-based models trained on our dataset achieve the best performance on both session search and click prediction tasks, surpassing BM25 and models trained on existing datasets in both Chinese and English settings. The results demonstrate the capacity of our dataset to assess various models and IR tasks. 
It can be seen that WARRIORS provides an ideal foundation for enhanced IR models within web search scenarios, containing near-real user interaction behaviors and the latest web search results.

%% file: sec/conclusions.tex
\section{Conclusion}

In this study, we introduced BASES, a new framework designed for the simulation of large-scale web search user behaviors with LLM-based agents. We designed specific strategies to construct large-scale user profiles, which can subsequently lead to human-like search behavior data in our framework.  
Through extensive experiments, we not only demonstrated the effectiveness of BASES in simulating authentic and diverse user behaviors, but also highlighted its potential in improving information retrieval tasks, especially in low-resource scenarios. 
Furthermore, to facilitate the related research, we develop the WARRIORS dataset, available in both Chinese and English versions, which comprises a large collection of simulated web search user behavior data by BASES. 
We believe that this work can provide new perspectives to investigate the search behaviors of web users, thereby contributing to the advancement of search technologies and user experience optimization.
For future work, we will enrich the behavior types of simulated users, and also consider applying the BASES framework to other fields like sociology analysis.

%% file: sec/appendix.tex
\section{Simulation Consistency Evaluation}
\label{appendix:consistency}
To demonstrate the feasibility of web search user simulation, we assess the consistency between the behavioral patterns exhibited by LLM agents and those of real-world users within the search engine.

\subsection{Settings}
\paratitle{Profile Generation.} We conduct the experiment on TREC Session~\citep{DBLP:conf/trec/CarteretteKHC14}, a human-annotated session search dataset encompassing user sessions of query-search-click behaviors, aimed at fulfilling informational needs of given task descriptions. Due to the limited volume of the dataset, we use ChatGPT as LLM-based agents to conduct real data based profile generation method in Section~\ref{sec:profile_methods} with existing task descriptions.

\paratitle{Evaluation Tasks.} Given the impracticality of evaluating the consistency of entire behavioral sequences, we deconstruct them and assess the consistency from two distinct behaviors: query and click. The query behavior evaluation including query generation and query rewriting. Query generation entails assessing the consistency between the initial queries generated by LLM agents, and those produced by humans. Query rewriting involves evaluating LLM agents' capabilities in formulating the subsequent query based on prior behaviors, ensuring it aligns with the human-rewritten query.
The click behavior evaluation focuses on the accuracy of click behaviors. It evaluates the consistency of LLM agents' selections in search results based on historical behaviors with those of humans.

\paratitle{Evaluation Metrics.}
We use two metrics to evaluate query behaviors: term overlap rate and GPT-4-based evaluation. The term overlap rate quantifies the extent of shared subject keywords between paired queries. This is determined by identifying common words after removing stop words, since shared terminology indicates thematic similarity and a consistent search intent. Despite its simplicity and efficacy, term overlap rate has limitations in overlooking synonymous terms and varying morphological forms. GPT-4-based evaluation is an supplement, which assesses the consistency of the query generated by the LLM agent with the human-query's search intent and its adherence to the user profile.
To evaluate click behaviors, we utilize top-1 accuracy to measure the proportion of instances where the LLM agent's clicked web page correspond to the same click of the human.

\subsection{Results and Analysis}
For query generation, the term overlap rate between queries generated by LLM agents and those produced by humans is found to be $90.8\%$, accompanied by the GPT-4 evaluation of $99\%$. For query rewriting, rewritten queries from LLM agents show a term overlap of $82.5\%$ with human-rewritten queries, accompanied by the GPT-4 evaluation of $98.8\%$. The findings suggest that the LLM agents are capable of generating queries that align closely with human search objectives, both in scenarios involving and lacking historical interactions.
Regarding the click prediction, the top-1 accuracy of the LLM agents' clicks mirroring human clicks is 51\%. Given the stringent nature of this metric, we manually review cases where the LLM agents do not precisely match the web pages clicked by humans. We find that in over 90\% of these cases, the web pages clicked by the LLM agents have minor discrepancy with those of humans and consistent with user profiles. This indicates that click behavior is influenced by the homogenization of search results and individual bias in human participants, and fundamentally, the LLM agent's click behavior aligns well with human patterns.

Overall, across both dimensions of query and click, LLM agents demonstrate a high degree of consistency with human behaviors, thereby ensuring the viability of user simulation.

\section{Prompt Design}
\label{appendix:prompts}
To prompt agents for executing precise actions, we devise two instruction strategies, including query behavior prompting and click behavior prompting, the specific prompt templates are shown in Table~\ref{tab:prompt_eng} for English setting and Table~\ref{tab:prompt_zh} for Chinese setting.

\section{Personalization Evaluation}
\label{appendix:personalization}

Whether user behaviors are personalized is a critical evaluation factor in user simulation, especially when simulating large-scale web search users. If the simulation lacks personalization, it can lead to homogenization in the behavior of simulated users, meaning similar users are likely to perform the same search actions, which is detrimental to subsequent tasks. 

Therefore, we focus on assessing the personalization in the simulated user behaviors of BASES framework. We construct a series of simulated user pairs, which demonstrate high similarity across multiple attributes in their profiles. For each pair of similar users, we manually analyze the behavior data generated by our BASES framework, the Cohen’s Kappa of human participants is 0.9.
We find that even users with extremely similar profiles (differing in only 1 or 2 attributes) exhibited significantly inconsistent behavior patterns. 
The evaluation result proves that the user behaviors simulated with BASES framework possess a significant degree of personalization. Furthermore, it also demonstrate that BASES framework provides deep insights into the complexity and diversity of user behavior patterns.

\begin{table*}[]
    \centering
    \small
    \begin{tabular}{p{15cm}}
        \toprule
        \textit{Query Behavior Prompting} \\
        \midrule
You are a search engine user with your own profile. Your task is to interact with search engines \{max\_exceeds\_times\} times. You have two types of operations to perform:

- Search[query]: When the current round has not reached \{max\_exceeds\_times\} times, please raise the next question based on your profile and web browsing history. Your query should be an entity phrase that has a similar topic to an attribute in your user profile. The query must be concise and clear. For example, Search[bollywood growth], Search[hjunk food trax], Search[pseudocyesis information], Search[location of port arthur].

- Finish[finish], When the current round exceeds \{max\_exceeds\_times\} times, you need to end your interaction with search engines. For example, Finish [Finish]
\vspace{0.5cm}

** Your Profile **
\{profile\}
\vspace{0.5cm}

** Web browsing (click) history **
\{scratchpad\}
\vspace{0.5cm}

** Your action ** \\
\toprule
\textit{Click Behavior Prompting} \\
\midrule
You are a search engine user with your own profile. Your task is to click on the most relevant page. 

In this interaction, you raised the question of **{query}**. You have received several webpage titles returned by the search engine. 

Now, based on your profile, web browsing history, and the relevance between the query and the titles, please choose the most relevant webpage to click on. Please note that you can only output one number from 1 to 10 to represent the title you are about to click on, and cannot output any other content
\vspace{0.5cm}

** Your Profile **

\{profile\}
\vspace{0.5cm}

** Web browsing (click) history **

\{scratchpad\}
\vspace{0.5cm}

** Query **

\{query\}
\vspace{0.5cm}

** Titles **

\{titles\}
\vspace{0.5cm}

** Your click **
\\ 
        \bottomrule
    \end{tabular}
    \caption{Various prompting strategies with query behavior prompting and click behavior prompting in English.}
    \label{tab:prompt_eng}
\end{table*}

\begin{table*}[]
    \centering
    \small
    \begin{tabular}{p{15cm}}
        \toprule
        \textit{Query Behavior Prompting} \\
        \midrule
\begin{CJK*}{UTF8}{gbsn}
您是一位带有自己档案的搜索引擎用户。您的任务是与搜索引擎互动\{max\_exceeds\_times\}次。您有两种操作可以执行：

- Search[查询]：在当前轮次未达到\{max\_exceeds\_times\}次时，请同时参考网页浏览历史和档案提出下一个查询。你需要考虑的是：1、在参考网页浏览历史时，查询的主题不能过多偏离历史中的交互内容，保持主题的连贯性；2、在参考档案时，查询的主题要和档案中的某属性相关，但要有发散思维，不要限制在这些具体属性上，尤其注意查询不要过多的涉及地名。

最后，您的查询应该是一个实体短语，保持简洁明了。这里有一些该操作的例子，如Search[优酷客户端下载], Search[无敌铁桥三], Search[蜡笔小新], Search[百度云网盘资源]。

- Finish[finish]：在当前轮次超过\{max\_exceeds\_times\}次时，您需要结束与搜索引擎的互动。例如，Finish[finish]。
\vspace{0.5cm}

** 您的档案 **

\{profile\}
\vspace{0.5cm}

** 网页浏览（点击）历史 **

\{scratchpad\}
\vspace{0.5cm}

** 您的操作 ** 
\end{CJK*} \\
        \toprule
        \textit{Click Behavior Prompting} \\
        \midrule
\begin{CJK*}{UTF8}{gbsn}
您是一位带有自己档案的搜索引擎用户。您的任务是点击最相关的页面。

在这次互动中，您提出了**{query}**的查询。您已经收到了搜索引擎返回的几个网页标题。

现在，请根据您的档案、网页浏览历史以及查询和标题之间的相关性，选择最相关的网页进行点击（语义相关或者是词语相似度高）。请注意，请首先输出你选择该title的理由，然后输出一个 **数字** （1到10）来代表您即将点击的标题，例如我会选择Title x: xxxx, 因为xxx。 
\vspace{0.5cm}

** 您的档案 **

\{profile\}
\vspace{0.5cm}

** 网页浏览（点击）历史 **

\{scratchpad\}
\vspace{0.5cm}

** 查询 **

\{query\}
\vspace{0.5cm}

** 标题 **

\{titles\}
\vspace{0.5cm}

** 您的点击 **
\end{CJK*} \\
        \bottomrule
    \end{tabular}
    \caption{Various prompting strategies with query behavior prompting and click behavior prompting in Chinese.}
    \label{tab:prompt_zh}
\end{table*}